\documentclass[letterpaper]{article} 
\usepackage[submission]{aaai24}  
\usepackage{times}  
\usepackage{helvet}  
\usepackage{courier}  
\usepackage[hyphens]{url}  
\usepackage{graphicx} 
\usepackage{natbib}  
\usepackage{caption} 
\usepackage{algorithm}
\usepackage{algorithmic}

\usepackage{newfloat}
\usepackage{listings}
\DeclareCaptionStyle{ruled}{labelfont=normalfont,labelsep=colon,strut=off} 
\floatstyle{ruled}
\newfloat{listing}{tb}{lst}{}
\floatname{listing}{Listing}
\usepackage{amsmath,amsfonts}
\usepackage{subfigure}

\usepackage{booktabs}       

\usepackage{pifont}

\usepackage{multirow}

\title{Can we only use guideline instead of shot in prompt?}
\author{
 \textbf{Jiaxiang Chen\textsuperscript{1}\thanks{This work is done during the student internship at Microsoft.}},
 \textbf{Song Wang\textsuperscript{2}},
 \textbf{Zhucong Li\textsuperscript{1}},
  \textbf{Wayne Xiong\textsuperscript{2}},
  \textbf{Lizhen Qu\textsuperscript{3}},
   \textbf{Zenglin Xu\textsuperscript{1}},
    \textbf{Yuan Qi \textsuperscript{1}}
\\
\textsuperscript{1}Fudan Univeristy,
\textsuperscript{2}Microsoft Azure AI,
\textsuperscript{3}Monash university
}

\usepackage{bibentry}

\begin{document}

\maketitle

\begin{abstract}
Currently, prompting techniques can be mainly divided into two categories:1) \textbf{shot} method implicitly inspires the model to answer the question by mimicing the steps in the given example, e.g., the few-shot CoT. 2) \textbf{Guideline} method explicitly instructs the model to reason by following guidelines, which contains succinct and concise task-specific knowledge.
Shot method is prone to difficulties in terms of selection of shots type, the number of shots, and the design of the reasoning steps, so a question arises: can we only use guideline instead of shot in the prompt?
To this end, we propose the \textbf{FGT} framework to automatically learn task-specific guidelines from dataset consisting of \textbf{F}eedback, \textbf{G}uideline, and \textbf{T}ree-gather agents.
First, the feedback agent is designed to evaluate the outcomes, both right and wrong, of each Q\&A to gather insights guiding more effective optimization strategies.
Next, the guideline agent is tasked with deriving guidelines from each piece of feedback and storing them in local memory. Lastly, the tree-gather agent aggregates all guidelines hierarchically through a tree structure,ultimately obtaining all unduplicated guidelines from a global perspective.
In addition, we induce the model to generate intermediate processes to ensure the reasoning consistent with the guidelines.
Experimental results demonstrate that our approach achieves superior performance across multiple tasks, thereby highlighting the effectiveness of using the guidelines in prompt.
\end{abstract}

\section{Introduction}

\label{introduction}
As LLM continue to advance~\cite{transformer,bert,t5,gpt3,gpt4,llama}, prompting techniques are becoming increasingly important due to the observation that well-crafted prompt can significantly improve the quality and correctness of responses~\cite{cot,got,tot,cot_sc,symbol_cot}.
\begin{figure}[tb]
	\centering
	\includegraphics[scale=0.9]{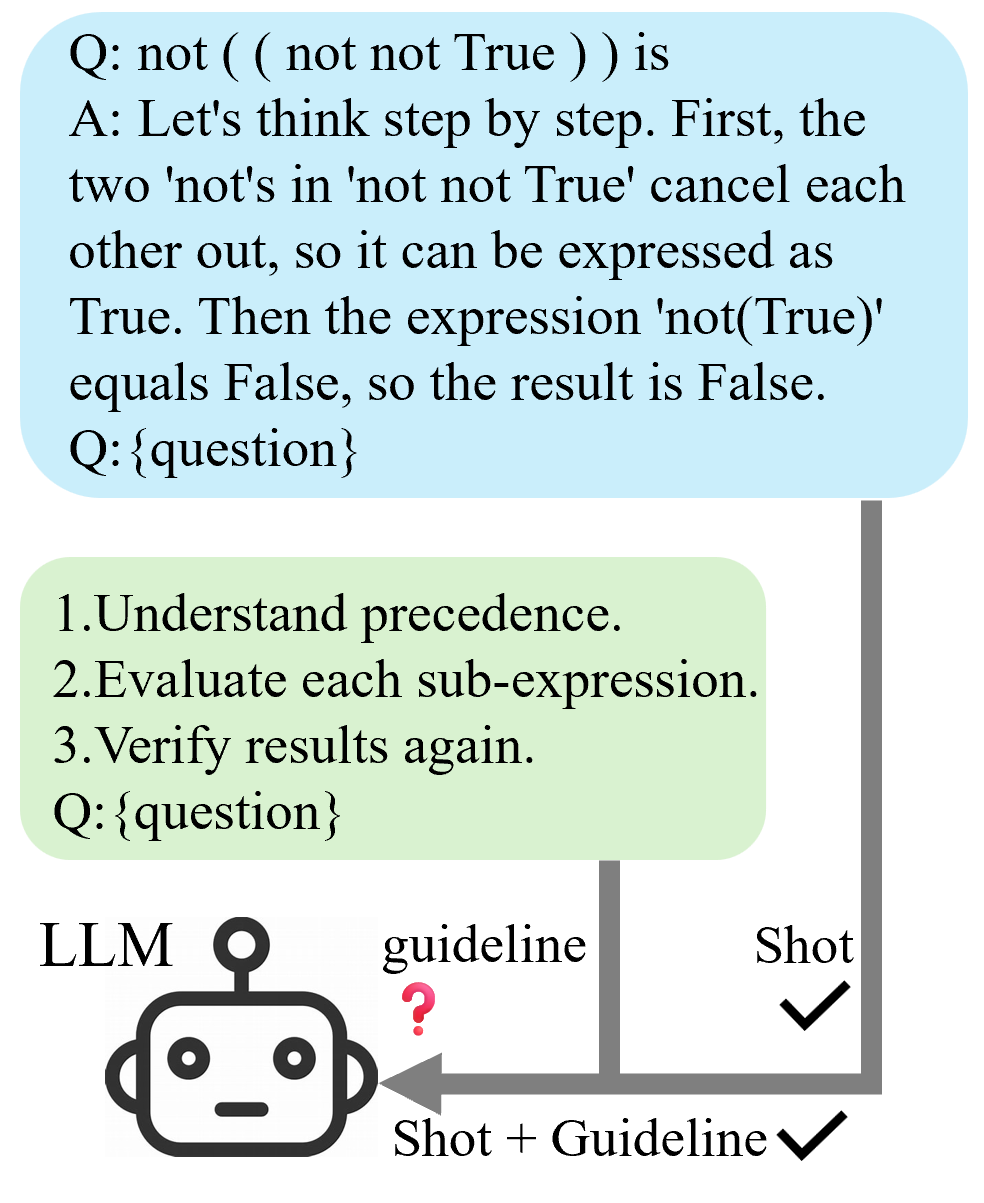}
	\caption{It's commonly using the shot or shot with guideline to assist the LLM to answer the question. Can we only use the guideline in prompt, to achieve comparable performance?}
	\label{fig:illustration_g_vs_shot}
\end{figure}

Currently, Techniques for assisting models in prompts generally fall into two categories: \textbf{shot} methods and \textbf{guideline} methods, as shown in Figure ~\ref{fig:illustration_g_vs_shot}.
\textbf{Shot} method implicitly inspires the model to answer the question by mimicing the reasoning steps in the selected examples, e.g., the few-shot CoT(the given shot is reasoned by using the CoT~\cite{cot}), which tends to have a longer input. \textbf{Guideline} method explicitly instructs the model to reason by following guidelines summarized from the task data. 

Most of the existing methods use human-crafted shot or shot combined with guideline to assist model inference in Fig~\ref{fig:illustration_g_vs_shot}. However, \textbf{Shot} method depends not only on the quality of the human-crafted reasoning steps but also on the diversity of shots.  In addition, people have observed that the similarity between the shots and the question to be answered is also of great importance. 
On the other hand, a limited number of examples in shot cannot contain all useful question types, which may miss some task-specific knowledge. Therefore, Shot method is prone to difficulties in terms of selection of shots type, the number of shots, and the design of the reasoning steps. so a question arises: can we only use guideline instead of shot in the prompt?

The \textbf{Guideline} method effectively bypasses the difficulties of selection and design in the shot by extracting and summarising guideline information from the overall case, which contain concise and clear task-specific knowledge. In addition, guidelines can not only be summarised from the case crafted by humans, but can also be easily obtained from llm(autoprompt)~\cite{ape,apo,opro,leap,wang2023promptagent,promst,intent_prompt}, and for simplicity we adopt llm to target this problem.

However, many autoprompt methods rely heavily on \textbf{optimization trajectory}~\cite{opro}, which may not utilize all cases in the training set to effectively find promising directions to improve the generated instructions.
Meanwhile, the lack of structured representations of guidelines may cause forgetfulness during the learning process and difficulty in gathering globally useful information. 
In addition, we observe that many of the responses contain only answers without a reasoning process, so it is questionable that there was any reasoning or thinking that followed the guideline closely.

To mitigate these problems, we propose a multi-agent \textbf{FGT} framework consisting of feedback agent, guideline agent and the tree-gather agent. Inspired by some works~\cite{opro,promst,wang2023promptagent}, we design the feedback agent to analyse from each Q\&A including wrong or correct cases to obtain the feedback as a more effective direction of optimization. Also, we adopt the way of gathering all useful guidelines from a global perspective without using the optimization trajectory. Specifically, we employ a guideline agent to extract guideline from the feedback of each Q\&A and store them in local memory.  Lastly, the tree-gather agent aggregates all guidelines extracted from all Q\&As hierarchically through a tree structure, ultimately obtaining all useful guidelines in final prompt. Besides, we incorporate the "Please give the process, not only the answer." into the prompt to make the LLM to output the intermediate process rather than merely presenting the final answers.

To evaluate our method, we conduct experiments on the Big-bench Hard (BBH) dataset across multiple tasks which we categorize into math calculating,logic reasoning and context understanding. Results demonstrate that the guideline-based prompt obtained from our framework~\textbf{FGT} attains remarkable performance, thereby highlighting the effectiveness of using the guidelines in prompt.

Our contributions can be listed as follows:
\begin{enumerate}
\item  We propose the~\textbf{FGT} framework consisting of feedback, guideline, and tree-gather agents to learn the guideline based on the feedback of each Q\&A and gather them hierarchically to obtain the final prompt from a global perspective.
\item  We conduct a lot of experiments to verify whether we can only use the guideline instead of the few-shot in prompt and eventually demonstrate the effectiveness of using the guidelines in prompt.
\item  we induce the model to generate intermediate processes to ensure the reasoning consistent with the guidelines, which fills the gap of only having answers without steps, and significantly improves accuracy when adhering to the guidelines.
\end{enumerate}

\section{Related Work}
\label{related-works}
\subsection{Prompting and In-context learning}
Prompting and In-context learning (ICL) are emergent ability of large language models (LLMs) \cite{wei2022emergent} and have become efficient learning paradigms for LLM. The remarkable in-context learning (ICL) ability of LLMs also leads to efficient few-shot learners that can generalize from few-shot input-label pairs \citep{brown2020language,sahoo2024systematic}. 
Our approach facilitates the learning of task-specific guidelines from the data, with a distinct emphasis on crafting an intermediate process that rigorously adheres to these guidelines. This adherence significantly enhances the LLM's comprehension of the task, ultimately leading to improved and superior performance.
 
\subsection{LLM based multi-agent system}
Large Language Models (LLMs) have recently shown remarkable potential in reaching a level of reasoning and planning capabilities comparable to humans. This ability exactly aligns with the expectations of humans for autonomous
agents. Based on the development of using one LLM as a single planning or decision-making agent, LLM-based multi-agent systems have achieved considerable progress in complex problem-solving and world simulation~\cite{guo2024large,autogen,ai2apps}.
Therefore, we propose a multi-agent framework to learn the task-specific guideline-of-thought to mitigate the autoprompt problem, which consists of three agents: feedback agent, guideline agent and the tree-gather agent.

\subsection{Automatic prompt improvement}
Due to the importance and difficulty of prompt engineering in adapting LLM in real application, researchers have proposed various approaches related to automatic prompt generation and prompt improvement.
APE~\cite{ape} firstly introduce LLM for generating instruction prompts in three steps: candidate, selection and resampling. APO~\cite{apo} applies the gradient descent in language space to obtain the optimized prompt. OPRO~\cite{opro} further take LLM as an optimizer and improve the performance through iterations. PROMST~\cite{promst} integrates human feedback and adopts the score model to select the prompt efficiently compared to evaluating on dataset directly. PromptAgent~\cite{wang2023promptagent} utilizes the form of agent to retrieve error, generate feedback and update the prompt. IPC~\cite{intent_prompt} jointly generates synthetic data of boundary use cases and optimizes the prompt. LEAP~\cite{leap} combine the principles learned from the mistake with the few-shot CoT to facilitate the generation of refined answers. 
However, it is not clear whether Leap's performance relies more on the common rules in the guideline or on the solution steps in the few-shot CoT.To investigate this, we propose the \textbf{FGT} framework to learn the guidelines from the data automatically, and we incorporate the process prompt to make the reason aligning with the guidelines. 
\section{Method}
\label{methods}

\begin{figure*}
	\centering
	\includegraphics[scale=0.7]{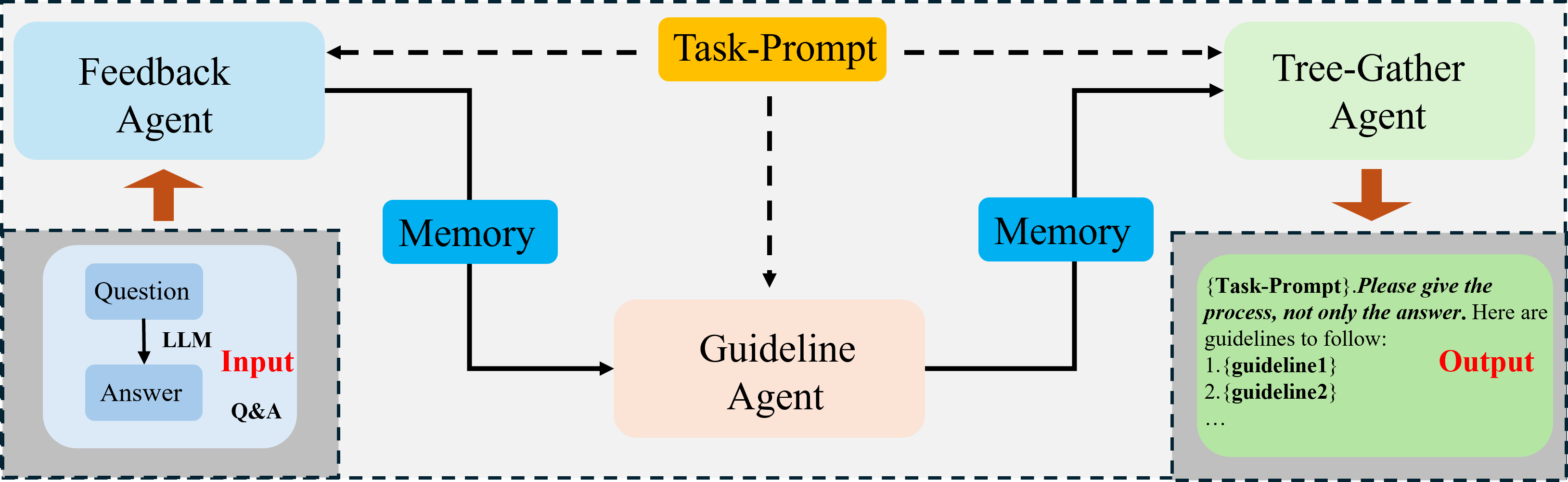}
	\caption{Overview of our proposed \textbf{FGT} framework for learning guidelines from the data,which consists of three agents:feedback agent,guideline agent and tree-gather agent. It takes the Q\&A pair as input, and generates the final prompts with guidelines, with the output of each step stored in memory for the next stage. Task-prompt describes the task in one sentence clearly and concisely.}
	\label{fig:framework}
\end{figure*}






To automatically extract and learn the guidelines from data, we propose the multi-agent framework FGT.  To make the model explicitly follow the guidelines, we explicitly add "please give the thought process, not the answer" in the final prompt.

As illustrated in the Fig~\ref{fig:framework}, Our multi-agent framework consists of three agents: the feedback agent, the guideline agent, and the tree-gather agent. They all take as input the task-prompt, which intends to clarify the task in one sentence concisely. 

The role of the feedback agent is to produce relevant feedback for each question and its corresponding answer provided by the LLM, based on the Input-Output (IO) process. It evaluates the accuracy of the question-answer duo by comparing it with the ground truth and then creates a detailed report that points out the specific aspects that should be considered for that task.

The guideline agent processes the question-answer combination and the feedback to create prompts that include general guidelines derived from each Q\&A instance. Additionally, if we have manual feedback, we can input it straight into the guideline agent without going through the feedback agent.

The purpose of the tree-gather agent is to compile and finalize the guidelines generated by the guideline agent. It organizes the information extracted from all the Q\&As into a hierarchical tree structure.

\subsection{Problem proposal}

We denote the question as \(x\), the answer as \(y\), the prompt with guidelines as \(g\), and the LLM as \(p()\). Therefore, the relationship can be formulated as \(y \sim p(y | x, g)\), indicating that the LLM takes the question \(x\) and the guideline-prompt \(g\) as input, resulting in the output answer \(y\).

Our goal is to develop a function $f$ to learn the prompt including final guidelines $g=f(x,y)$, so we divide them into three parts $f_{feedback}$, $f_{guideline}$ and $f_{gather}$ for the purpose of generating feedback, yielding the Q\&A-level guideline, and gathering guidelines separately, which can be formulated as below, where the $d$ means the feedback, the $i$ indicates the order and n is the length of the learned data:

\begin{equation}
\begin{aligned}
d_i=f_{feedback}(x_i,y_i)
\label{formulat:f1}
\end{aligned}
\end{equation}

\begin{equation}
\begin{aligned}
g_i=f_{guideline}(x_i,y_i,d_i)
\label{formulat:f2}
\end{aligned}
\end{equation}

\begin{equation}
\begin{aligned}
g=f_{gather}(g_1,g_2,...,g_n)
\label{formulat:f3}
\end{aligned}
\end{equation}

We propose three agents to implement the three functions described above, as shown in Fig~\ref{fig:framework}. The feedback agent generates feedback for the learner, the guideline agent produces the guideline for the feedback, and the tree-gather agent collects the guidelines using a tree structure.

\subsection{Feedback agent}
\begin{figure}[h]
	\centering
	\includegraphics[scale=0.55]{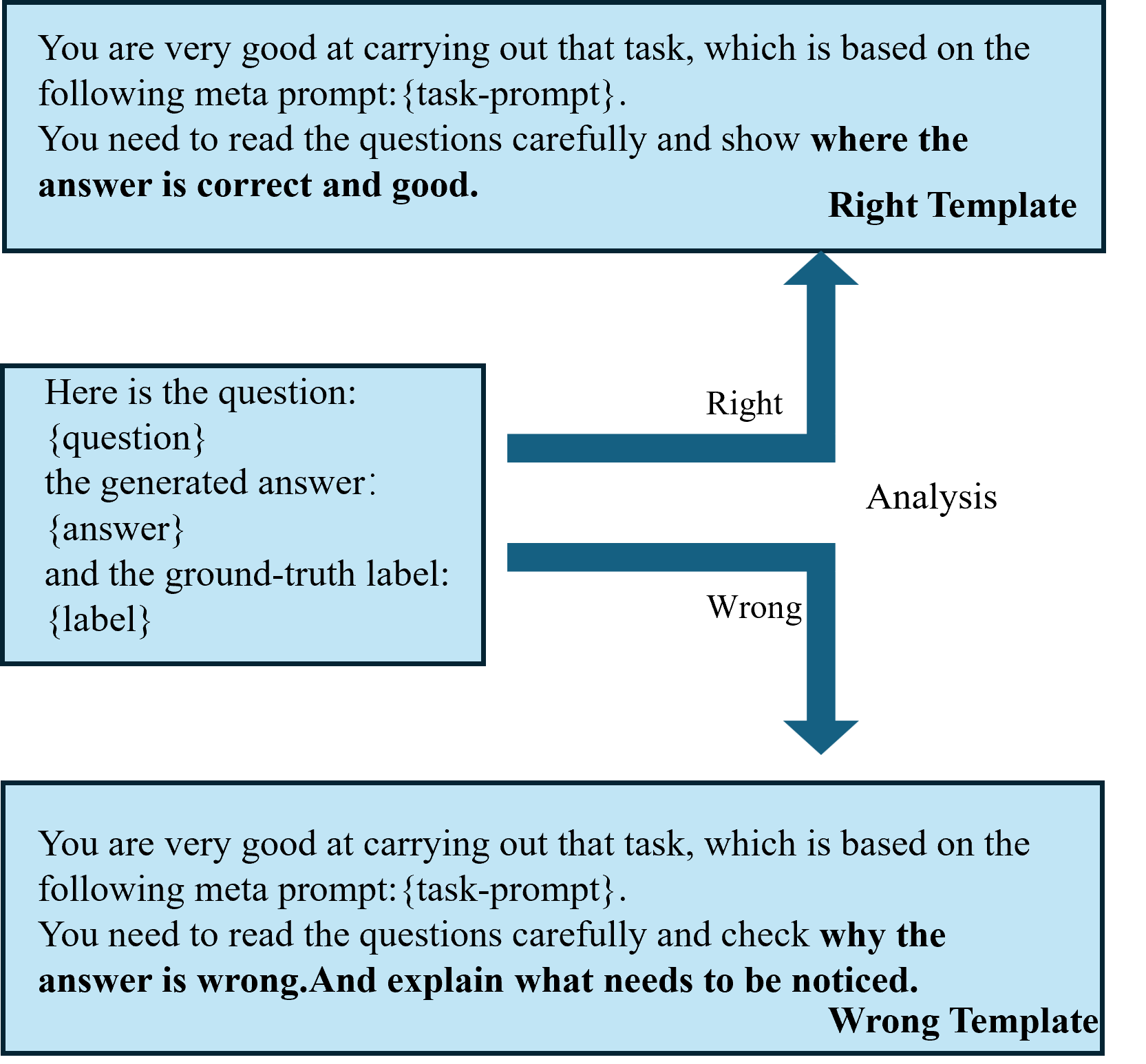}
	\caption{Details of our feedback agent, which includes the judgement and the analysis steps.}
	\label{fig:f_agent}
\end{figure}

As illustrated in the Fig~\ref{fig:framework} and Fig~\ref{fig:f_agent}, Our proposed feedback agent takes each $Q\&A$ pair including the question $x_i$ and the corresponding answer $y_i$ generated from the LLm, along with the task-prompt as input, generates the feedback $d_i=f_{feedback}(x_i,y_i)$as mentioned in the equation~\ref{formulat:f1} and stores all feedback in memory for use by the guideline agent.

Since LLMs tend to perform better on tasks that are more specific and less vague, we split this task into two sub-tasks: judgement and analysis. The judgement sub-task evaluates the correctness of the generated answer by comparing it with the ground truth label.

The analysis sub-task uses the task-prompt to generate more relevant feedback by examining the case according to the outcome of the judgement. The feedback is either positive or negative, depending on whether the answer is correct or incorrect. All feedback is generated by the AI without human involvement. However, if there is any human feedback available, we can directly input it to the guideline agent without using the feedback agent.
\subsection{Guideline agent}
\begin{figure}[h]
	\centering
	\includegraphics[scale=0.6]{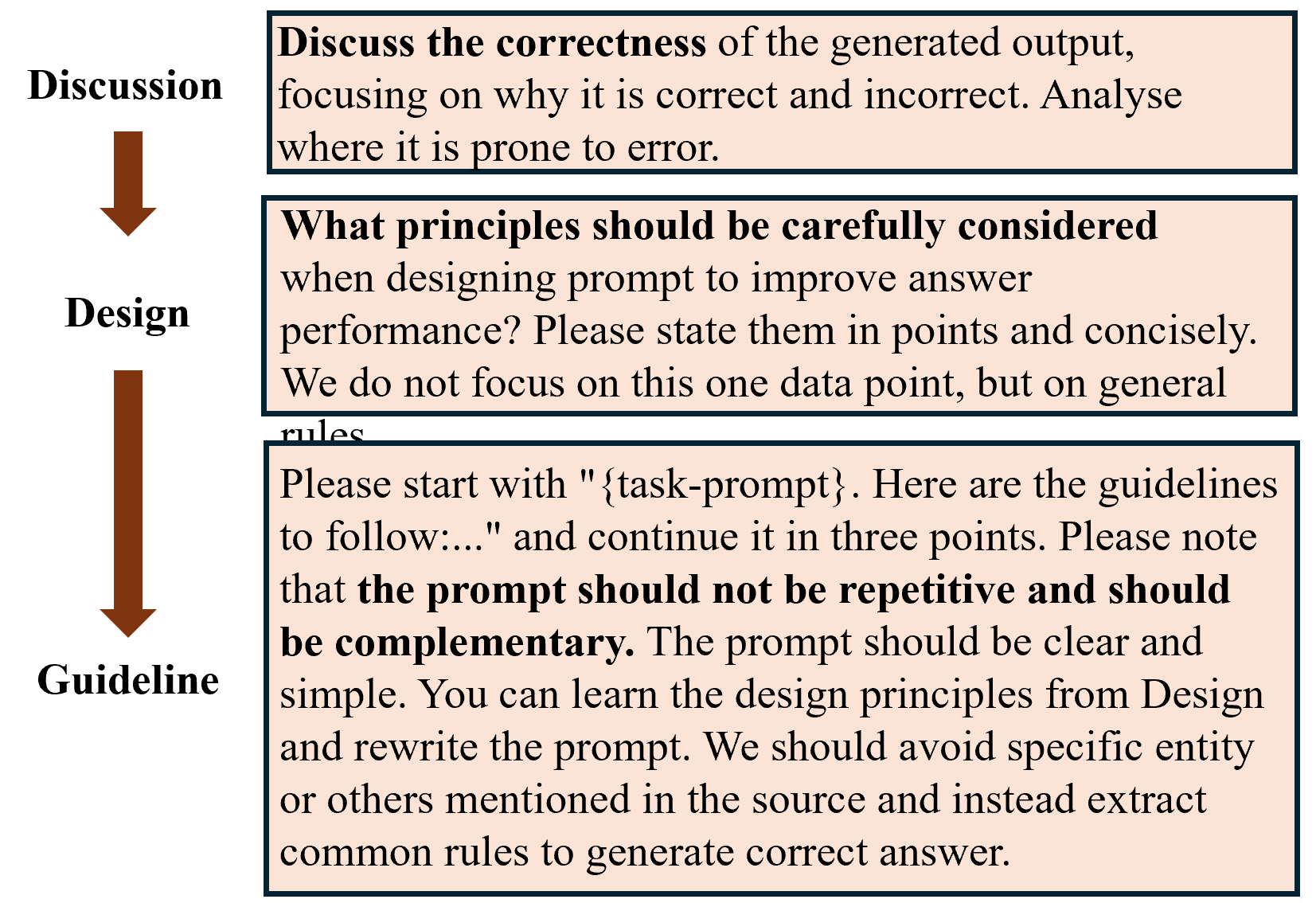}
	\caption{Details of our guideline agent,which involves the discussion,design and the guideline parts.}
	\label{fig:p_agent}
\end{figure}

We have formulated an architecture that includes discussion, design, and guideline generation. 
This takes advantage of large language models (LLMs) to perform specific and straightforward tasks as part of a more comprehensive process of step-by-step reasoning.


Figure~\ref{fig:p_agent} illustrates the three key stages within our guideline agent: discuss, design, and prompt. The inputs for this agent are not just the question and answer pair $(x_i,y_i)$ but also feedback $d_i$ in the memory from the feedback agent, as expressed in equation~\ref{formulat:f2}.

During the \textbf{Discusssion} process, we delve into why an answer is right or wrong through an analysis by the LLM, focusing on identifying potential errors informed by feedback related to both correct and incorrect responses.

In the \textbf{Design} phase, the goal is to develop overarching principles that will inform the guideline generation. We give special attention to learning broad, task-specific guidelines that prevent the formulation of overly specific principles.

Lastly, the \textbf{Guideline generation} stage establishes a prompt that incorporates these guidelines. The format for this final step is as follows:
\noindent
\textit{$<$task-prompt$>$. Please give the process, not only the answer. Here are the guidelines to follow:\\ $<$list of newly learned principles align with principle from design step$>$}.

\subsection{Tree-Gather agent}
\begin{figure}[htbp]
	\centering
	\includegraphics[scale=0.55]{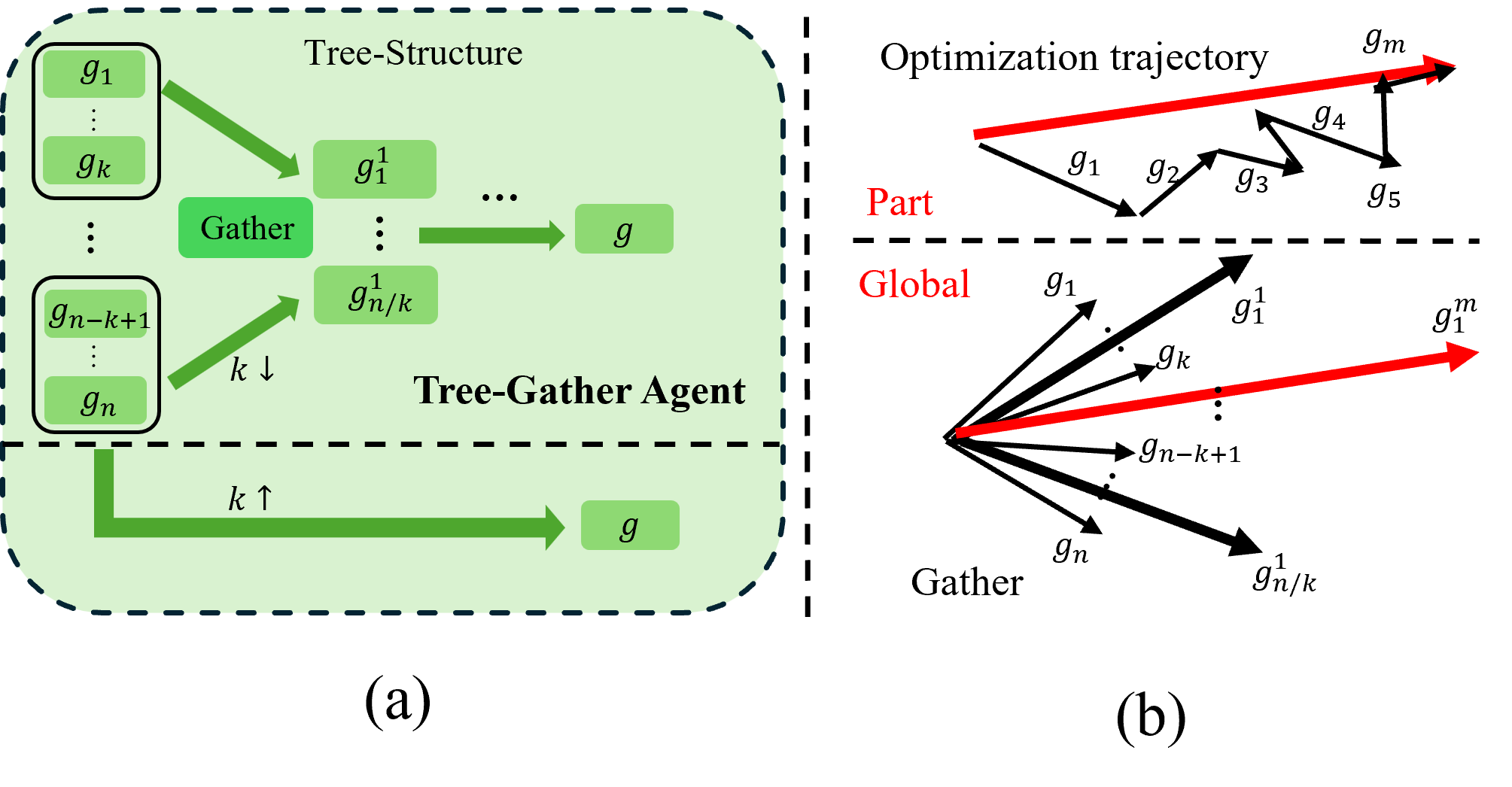}
	\caption{Illustration of our proposed tree-gather agent.(a) shows the gathering process using the tree structure, where it can be varied to one step of direct combining as k increases. (b) compares the way of gathering and the optimization trajectory from an optimization perspective, reflecting the global view of our approach.}
	\label{fig:t_agent}
\end{figure}

The goal of our tree-gather agent is to generate a final prompt $g$ with all useful guidelines which covers the full range of key points of the task from the data.  As shown in the figure~\ref{fig:t_agent}, unlike \textbf{optimization trajectory}~\cite{opro} which adjusts based on local information each time, we globally aggregate the guidelines learned from each Q\&A sample $g_i$, thus aggregating the all useful guidelines.

Considering the limited context window of LLM, we proposes tree-gather agent, which can regulate the complexity by adjusting k. And this paradigm adapts well to the context understanding capability of different LLMs. Like many applications of tree-structure in RAG~\cite{t-rag,raptor} to capture more detail of the source document, we believe that our proposed tree-gather agent is more applicable to the case of larger amount of data in the memory, if when the amount of data is smaller direct combine is sufficient.

Algorithm~\ref{alg:algorithm} shows the details of our tree-gather agent.
We first perform a sliding window aggregation of size k on the data without overlapping.
Then, we repeat the same "k" operations on the aggregated guidelines, until there is only one result. This is the final guideline.
For simplicity, we set $k=5$ throughout the experiments.

\begin{algorithm}[t]
\caption{Learning guidelines from data through our proposed multi-agent \textbf{FGT} framework.}
\label{alg:algorithm}
\begin{algorithmic}[1]
\REQUIRE Question $x=\{x_1,x_2,...,x_n\}$, answer $y=\{y_1,y_2,...,y_n\}$ and Arity of tree $k$.
\FOR{$i=1:n$}   
        \STATE $d_i=f_{feedback}(x_i,y_i)$   
        
        \STATE $g_i=f_{guideline}(x_i,y_i,d_i)$   
    \ENDFOR
    \WHILE{n$>$1}                 
        \STATE $j=1$
        \FOR{$i=1:\lceil \frac{n}{k} \rceil$}  
        \STATE $g_i=f_{gather}(g_j,g_{j+1},...,g_{j+k-1})$
        \STATE $j=j+k$ 
        \ENDFOR
        \STATE $n=i$ 
    
    \ENDWHILE
\ENSURE g  
\end{algorithmic}
\end{algorithm}

\section{Experiment}
\subsection{Dataset}
\begin{figure*}[htbp]
	\centering
	\includegraphics[scale=0.55]{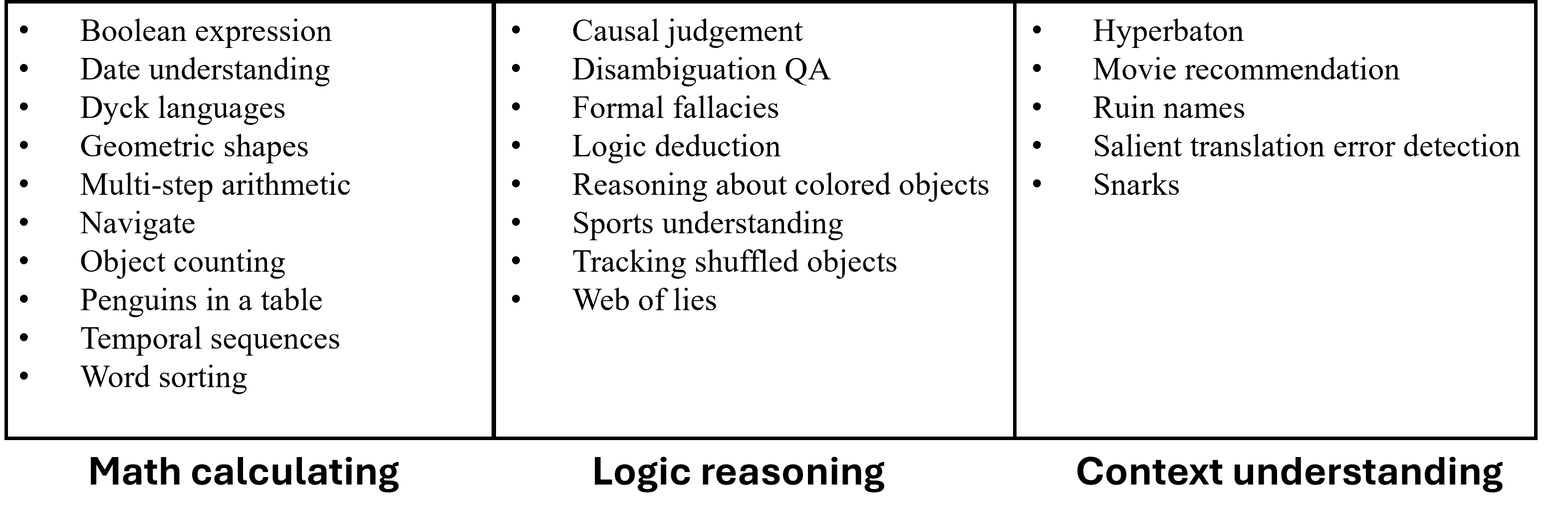}
	\caption{The re-categorized bbh dataset.It includes three main task:Math calculating, logical reasoning and in context understanding.}
	\label{fig:bbh_3_task}
\end{figure*}
Big-bench-hard(BBH) is a subset of BIG-Bench and consists of 23 challenging tasks, among them the logical deduction and tracking shuffled objects each contain three sub-tasks, making a total of 27 tasks.

We categorize these tasks into three main types: math calculating, logical reasoning, and context understanding, as shown in Fig~\ref{fig:bbh_3_task}.

\textbf{Math calculating} relates to the Boolean operations, number calculations, date calculations, coordinate calculations, geometric  calculations, alphabetical ordering and other tasks involving mathematical symbolic calculations.

\textbf{Logical reasoning} involves the tasks that require a combination of logical thinking skills such as causal judgements, referential speculation, logical judgements, and substantive inference.

\textbf{Context understanding} contains sentence understanding across different languages, word understanding, content recommendations tasks, all of which heavily rely on the comprehension of the context.

\subsubsection{Baseline}
\noindent\textbf{IO (Input-Output).} IO takes only the task-prompt and question to provide an answer, useful for gauging LLM performance and benchmarking against other methods.

\noindent\textbf{CoT (Chain-of-Thought).} CoT prompts the LLM to think about the question step by step, by using the prompt "Let's think step by step," proven effective in various tasks.

\noindent\textbf{Few-shot.} Few-shot method employs a handful of $Q\&A$ samples ain the prompt, guiding the LLM by analogizing current questions to provided samples. Our study uses three examples.

\noindent\textbf{Few-shot-CoT.} Integrating few-shot and CoT, this approach provides demonstration examples coupled with CoT reasoning to steer the LLM toward thought-based answers, using the same samples as the few-shot experiment.

\noindent\textbf{Many-shot.} Many-shot~\cite{manyshot} shows LLM can achieve better results by simply including many (up to thousands) demonstration examples in the prompt. Following the same experiment description, we directly carry out the evaluation by including all training $Q\&A$ data examples in the input prompt.

\noindent\textbf{Leap (learning principles).} Leap~\cite{leap} intentionally induce the model to make mistakes and learn the principle from them, and test on the unseen examples by combining the principles and the crafted few-shot examples. 

\subsection{Experiment settings}
\subsubsection{Setting}

For our experiments, we random select 25\% of the data related to each task as the training set and the rest 75\% as the test set. 

The LLM used in all experiments in this paper is GPT-4-32k-0613 endpoint with the $temprature=0.7$ and the $top\_p=0.95$.
\subsubsection{Metric}
We use the accuracy as the metric,which is formulated as below, $N$ is the number of data: 
\begin{equation}
    \texttt{Accuracy} = \frac{\sum_{i=1}^{N} \mathbf{I}(y_{pred}==y_{true})}{N}
\end{equation}

\subsection{Main results}\label{main-results}

\begin{table}[t]
	\renewcommand{\arraystretch}{1.1}
	\small
	\setlength{\tabcolsep}{2mm}
	\begin{center}

        \resizebox{\columnwidth}{!}{
		\begin{tabular}{|c|c|c|c|c|c|}
        \hline
 	{Task} & IO & CoT   & Few-shot CoT  & Leap & Ours\\
		\hline
        Math calculating & 0.681 & 0.738  & 0.876  & 0.883 & \textbf{0.895}\\
        Logic reasoning & 0.669 & 0.859  & 0.884  &  0.891 & \textbf{0.939}\\
        Context understanding & 0.788 & 0.787  & 0.873 & 0.873 & \textbf{0.881}\\
		\hline
		\end{tabular}}
	\end{center}
\caption{Quantitative Comparison with Chain-of-Thought.}
\label{tab:quanti_cot}
\end{table}

\begin{table}[t]
	\renewcommand{\arraystretch}{1.1}
	\small
	\setlength{\tabcolsep}{2mm}
	\begin{center}

		\begin{tabular}{|c|c|c|c|}
        \hline
 	{Task}   & Few-shot  & Many-shot & Ours\\
		\hline
        Math calculating  & 0.675  & 0.724 & \textbf{0.895}\\
        Logic reasoning  & 0.674  & 0.657 & \textbf{0.939}\\
        Context understanding  & 0.823  & \textbf{0.884} & 0.881\\
		\hline
		\end{tabular}
	\end{center}
\caption{Quantitative Comparison with few-shot and many-shot.}
\label{tab:quanti_shot}
\end{table}

As shown in Table~\ref{tab:quanti_cot} and Table~\ref{tab:quanti_shot}, we present accuracy results for three groups of tasks: Math calculating, Logical reasoning, and Context understanding. 
These results are summarized from the average accuracy rates for the corresponding sub-tasks illustrated in Fig~\ref{fig:bbh_3_task}. Detailed data for each category can be found in Table \ref{tab:quanti_cot_detail} and Table \ref{tab:quanti_shot_detail} provided in the Appendix.

\subsubsection{Comparison with Variants of CoT}
In this section, we compare our result with the guideline-based method involving chain-of-Thought process. The data shown in Table~\ref{tab:quanti_cot} corroborates our analysis, indicating that our methodology outperforms others related to the Chain-of-Thought approach.





Our technique achieves a notable advancement in logic reasoning tasks, thereby underscoring its exceptional capability in handling such tasks.


Overall, compared to heuristic CoT methods without knowledge integration, manually crafted few-shot CoT, and improved few-shot CoT through guiding principles, our method demonstrates greater efficacy, confirming the effectiveness of incorporating guidelines into the prompts.
\subsubsection{Comparison with Many-shot}

Our approach is to learn the guideline from the training samples, so this section examines how effective it is to put the same all samples directly in the prompt without extracting and summarizing information like many-shot~\cite{manyshot}. Meanwhile we investigates the influence about the performance disparity between the few-shot and the many-shot.

According to Table~\ref{tab:quanti_shot}, our technique typically surpasses both few-shot and many-shot methods. Despite expectations from~\cite{manyshot} that many-shot should be superior to few-shot, it falls short in logic reasoning tasks, indicating that an abundance of examples doesn't ensure logical reasoning proficiency, thus underscoring the necessity for effective guidelines.

In tasks requiring context comprehension, many-shot excels above few-shot and our approach, pointing to such tasks' reliance on the LLM's innate contextual grasp, learned through exposure to many examples.

Overall, our method is most effective in mathematical calculation and logic reasoning, while many-shot prevails in context understanding tasks due to the LLM's intrinsic contextual insights.
\subsubsection{Comparing with Autoprompt}
\begin{table*}[t]
	\renewcommand{\arraystretch}{1}
	\small
	\setlength{\tabcolsep}{4mm}
	\begin{center}
	
		\begin{tabular}{|c|c|c|c|c|c|}
        \hline
        Method        & Penguins & Geometry & Object Count & Temporal & Causal Judge \\
        \hline
        APE           & 0.848    & 0.445    & 0.852        & 0.992    & 0.740       \\
        PromptAgent   & 0.962    & 0.680    & 0.888        & 0.982    & 0.770  \\
        \hline
        FGT(ours)          & \textbf{0.963}    & \textbf{0.705}    & \textbf{0.989}        & \textbf{0.995}    & \textbf{0.771}\\
		\hline
		\end{tabular}
	\end{center}
\caption{Quantitative Comparison with some autoprompt methods.}
\label{tab:autoprompt}
\end{table*}

We also select five tasks from BBH to verify the performance between APE, PromptAgent and our FGT. Both methods update prompt by iteratively using partial samples without a global update perspective. 

As seen in the Table~\ref{tab:autoprompt},we can see our FGT method obtains the best performance across these tasks, which also demonstrates the superiority of learning from a global perspective.

\subsection{Ablation study}\label{ablation-study}
\subsubsection{Incorporating a Process Prompt}
\begin{table}[t]
	\renewcommand{\arraystretch}{1.1}
	\small
	\setlength{\tabcolsep}{2mm}
	\begin{center}
        \resizebox{\columnwidth}{!}{
		\begin{tabular}{|c|c|c|}
        \hline
 	{Task}   & w/o add  & Ours\\
		\hline
        Math calculating  & 0.810  &\textbf{0.895}\\
        Logic reasoning  & 0.897  & \textbf{0.939}\\
        Context understanding  & 0.872  &\textbf{0.881}\\
		\hline
		\end{tabular}}
	\end{center}
 	\caption{Ablation study: Accuracy Comparison between whether adding the "\textbf{Please give the process, not only the answer.}" in the prompt.}
	\label{tab:abl_process}
\end{table}
\begin{figure*}[htbp]
	\centering
	\includegraphics[scale=0.6]{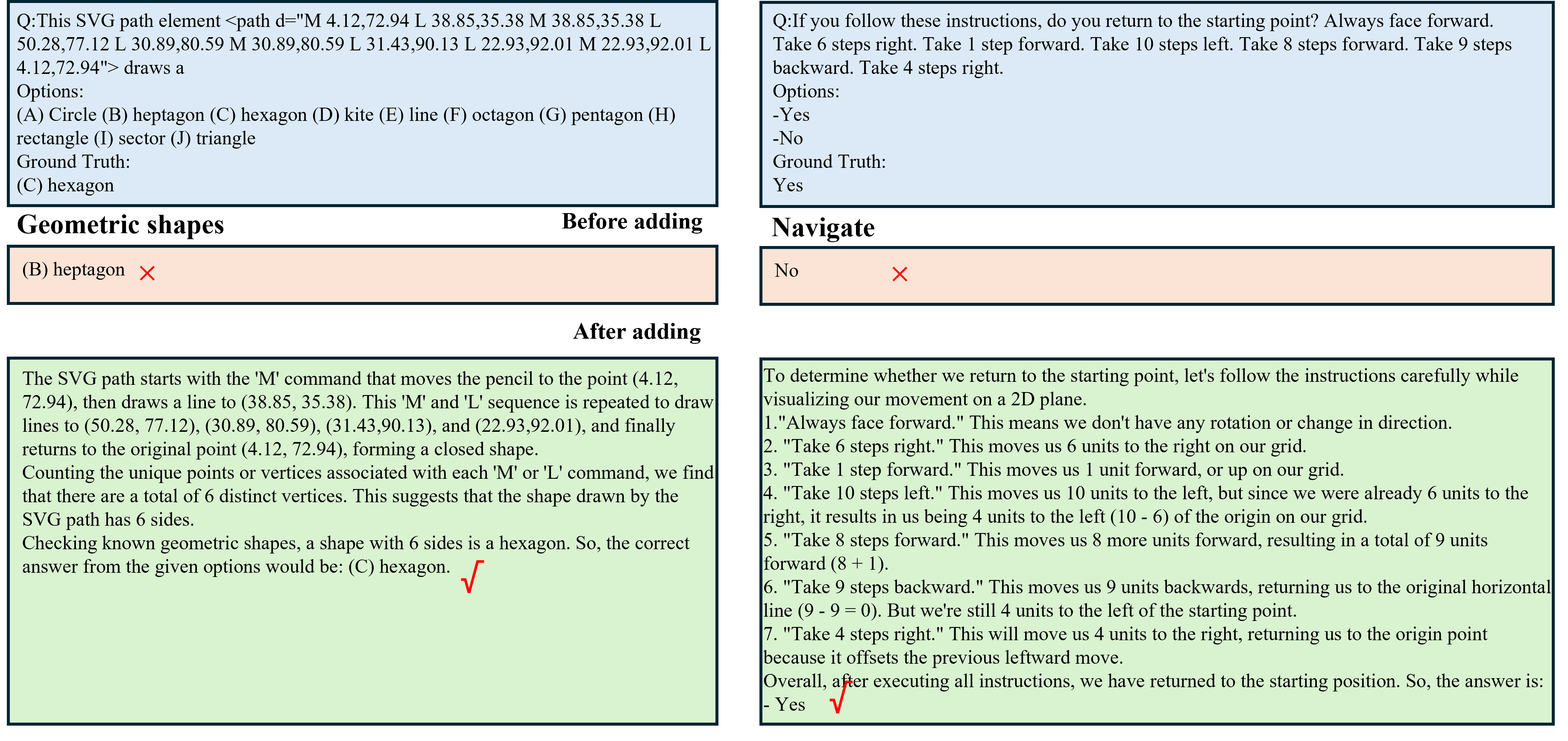}
	\caption{Illustration of the comparison between whether adding the prompt "\textbf{Please give the process, not only the answer.}".}
	\label{fig:cmp_sp}

\end{figure*}

This section examines the effects of using the process prompt "\textbf{Please give the process, not only the answer.}"

As illustrated in Fig~\ref{fig:cmp_sp}, the LLM's output tends to provide direct answers, bypassing the thought process without this specific instruction. This suggests that the LLM does not adequately apply learned guidelines, leading to incorrect responses.

On the other hand, incorporating the directive "\textbf{Please give the process, not only the answer.}" into the prompt encourages a comprehensive reasoning process that adheres to the guidelines, an essential step for accurate outcomes.

Additionally, accuracy results in Table~\ref{tab:abl_process} show a noticeable decrease performance in mathematical calculation and logical reasoning tasks without the process prompt, alongside a minor reduction in contextual comprehension tasks. This correlates with prior analysis indicating that these tasks depend heavily on the intrinsic abilities of the LLM.

\subsubsection{Scoring prompts using LLM.}
\begin{table}[t]
	\renewcommand{\arraystretch}{1.1}
	\small
	\setlength{\tabcolsep}{4mm}
	\begin{center}
		\label{tab:abl_score_prompts}
  
		\begin{tabular}{|c|c|}
        \hline
 	Gather method    & mean↑ \\ \hline
        Before gather &83.97     \\ \hline
        Direct combine  &89.74       \\ \hline
         Gather level 1 & 88.60  \\ \hline
         Gather level 2 & 89.21  \\ \hline
         Gather level 3 & \textbf{89.78} \\ \hline
		\end{tabular}
  	\caption{Score the prompts. Here shows the averaged accuracy result across all tasks.}
	\end{center}
\end{table}
Inspired by the score model in the PROMST~\cite{promst} instead of using the test accuracy~\cite{apo,opro}, we suppose it is also feasible to leverage the LLM's ability to design the prompts to score the prompt with learned guidelines. 
Due to the high cost of direct evaluating many prompts gathered by the tree-gather agent, we also adopt the score scheme to help investigate the gain of the tree-gather form.

We have designed five scoring criteria:1) adhering to the task, 2) generality, 3) comprehensive, 4) logical relationship, 5) correctness and accuracy.

\textbf{Verify Correlation}
First, we rated all prompts generated from our framework in each task. Then, for cost considerations, we sorted them by rating from lowest to highest and divided them into three groups. Subsequently, we randomly selected a prompt from these three groups and evaluated its performance on the test data.
As the Fig~\ref{fig:score_bar} shows, we can see the increasing trend which means the correlation is positive in almost all tasks in. 

\textbf{Investigate the tree-gather gain}
As illustrated in the Table~\ref{tab:abl_score_prompts}, we can see the average score performance across all task of the BBH. Among them, Before gather means the prompt with guidelines of each Q\&A, Direct combine meaning that directly feed all the prompt mentioned above into the LLM to obtain the final prompt with guidelines of all train dataset, which is the simplest method. And Tree-gather method gather all guideline of Q\&A hierarchically through a tree structure, where the level means the height of the tree, the level 3 is the final gathered prompt. The result shows that the score guideline of each Q\&A is the lowest, which align with the intutive that it only contain 

The results show that prompt with guideline extracted from each Q\&A has the lowest score, which is consistent with our intuition that it learns only very small features of the data and therefore generalises poorly. At the same time, the average performance of prompt with guideline continues to improve as the gather level increases, which also suggests that as the amount of data in the receptive field increases, the detailed guideline extracted from more data performs better. Meanwhile, the performance of directly combining all the data is only second to level 3 of tree-gather, which also verifies that increasing the amount of data and learning more task-specific guideline is beneficial to improve the performance of responses.

\section{Conclusion}
\label{conclusion}
In this paper, Can we only use guideline instead of shot in prompt? To investigate , we design a multi-agent \textbf{FGT} framework to automatically learn the task-specific guidelines from the data, which consists of three agents:feedback agent, guideline agent and tree-based gather agent.
The feedback agent analyse from each Q\&A including wrong or correct cases to obtain the feedback for a more effective direction of optimization. Also, instead of learning from optimization trajectory, we adopts the way of gathering all useful guidelines from a global perspective by our proposed guideline agent and the tree-gather agent. Besides, we incorporate the "Please give the process, not only the answer."in the prompt to make the LLM to reason align with the guidelines closely.
Compared with other Guideline methods like CoT,Leap and the Shot methods like Few-shot CoT, Experiment demonstrates the effectiveness of using the guidelines learned from our approach.

\bibliography{aaai24}

\clearpage
\appendix

\section{Appendix}
\label{sec:appendix}
\subsection{1.Full Comparison results on BBH with Chain-of-Thought.}

\begin{table*}[htbp!]
	\renewcommand{\arraystretch}{1.1}
	\setlength{\tabcolsep}{2mm}
	\begin{center}
		\begin{tabular}{|c|c|c|c|c|c|}
        \hline
 	dataset &IO & CoT   & Few-shot CoT  &Leap & Ours\\
		\hline
Boolean expressions	& 0.877&	0.978&	0.973& 0.976 &	1 \\
Causal judgement	&0.714	&0.75&	0.742& 0.727&	0.771 \\
Date understanding	&0.732	&0.767	&0.914	& 0.9 &0.963 \\
Disambiguation qa	&0.802	&0.828&	0.807& 0.852 &	0.973 \\
Dyck languages	&0.796	&0.834	&0.786	& 0.54&0.775 \\
Formal fallacies	&0.673	&0.754	&0.834	& 0.816 &0.84 \\
Geometricshapes	&0.439	&0.395	&0.518	&0.588 &0.705 \\
hyperbaton	&0.77	&0.75	&0.967	&0.988 &0.979 \\
Logical deduction three objects	&0.952	&0.936	&0.968	&1 &1 \\
Logical deduction five objects	&0.652	&0.716	&0.834	&0.844 &0.925 \\
Logical deduction seven objects	&0.668	&0.668	&0.556	&0.596 &0.866 \\
Movie recommendation	&0.791	&0.796	&0.887	&0.92 &0.995 \\
Multistep arithmetic two	&0.176	&0.872	&0.839	&0.924 &0.909 \\
navigate	&0.748	&0.684	&0.967	&0.984 &0.925 \\
Object counting	&0.679	&0.577	&0.973	&0.992 &0.989 \\
Penguins in a table	&0.715	&0.779	&0.963	&0.966 &0.963 \\
Ruin names	&0.887	&0.898	&0.925	&0.876 &0.92 \\
Sports understanding	&0.952	&0.941	&0.936	&0.944 &0.991 \\
Snarks	&0.797	&0.82	&0.909	&0.910&0.8 \\
Temporal sequences	&0.978	&0.967	&1	&1 &0.995 \\
Reasoning about colored objects	&0.877	&0.877	&0.935	&0.94 &0.983 \\
Salient translation error detection	&0.695	&0.673	&0.679	&0.672 &0.711 \\
Word sorting	&0.711	&0.524	&0.828	&0.956 &0.722 \\
Web of lies	&0.641	&0.86	&1	&1 &0.914 \\
Tracking shuffled objects seven objects	&0.395	&0.989	&1	&0.988 &1 \\
Tracking shuffled objects three objects	&0.369	&0.989	&1	&1 &1 \\
Tracking shuffled objects five objects	&0.337	&0.995	&1	&0.988 &1 \\
\hline
AVG	&0.697	&0.801	&0.879	&0.885 &0.912 \\
		\hline
		\end{tabular}
	\end{center}
\caption{Detailed Quantitative Comparison with Chain-of-Thought. Note that the Leap result is cited from \cite{leap}.}
\label{tab:quanti_cot_detail}
\end{table*}

\subsection{2.Full Comparison Results on BBH with Many-shot.}

\begin{table*}[htbp!]

	\renewcommand{\arraystretch}{1.1}
	\large
	\setlength{\tabcolsep}{2mm}
	\begin{center}

		\begin{tabular}{|c|c|c|c|}
        \hline
 	Dataset   & Few-shot  & Many-shot & Ours\\
		\hline
Boolean expressions	&0.877	&0.925	&1 \\
Causal judgement	&0.728	&0.75	&0.771 \\
Date understanding	&0.78	&0.79	&0.963  \\
Disambiguation qa	&0.786	&0.81	&0.973 \\
Dyck languages	&0.689	&0.737	&0.775 \\
Formal fallacies	&0.823	&0.786	&0.84 \\
Geometric shapes	&0.401	&0.652	&0.705 \\
Hyperbaton	&0.73	&0.946	&0.979 \\
Logical deduction three objects	&0.918	&0.877	&1 \\
Logical deduction five objects	&0.593	&0.657	&0.925 \\
Logical deduction seven objects	&0.561	&0.599	&0.866 \\
Movie recommendation	&0.887	&0.973	&0.995 \\
Multistep arithmetic two	&0.04	&0.059	&0.909 \\
Navigate	&0.73	&0.797	&0.925 \\
Object counting	&0.62	&0.7	&0.989 \\
Penguins in a table	&0.816	&0.752	&0.963 \\
Ruin names	&0.909	&0.92	&0.92 \\
Sports understanding	&0.898	&0.93	&0.991 \\
Snarks	&0.879	&0.872	&0.8 \\
Temporal sequences	&1	&1	&0.995 \\
Reasoning about colored objects	&0.946	&0.888	&0.983 \\
Salient translation error detection	&0.711	&0.711	&0.711 \\
Word\_sorting	&0.796	&0.824	&0.722 \\
Web of lies	&0.855	&0.61	&0.914 \\
Tracking shuffled objects seven objects	&0.241	&0.326	&1 \\
Tracking shuffled objects three objects	&0.39	&0.364	&1 \\
Tracking shuffled objects five objects	&0.348	&0.289	&1 \\
\hline
AVG	&0.702	&0.724	&0.912 \\
		\hline
		\end{tabular}
	\end{center}
\caption{Detailed Quantitative Comparison with few-shot and many-shot.}
  \label{tab:quanti_shot_detail}
\end{table*}
\subsection{3.Result of correlation between the score and the test accuracy}
\begin{figure*}[htbp]
	\centering
	\includegraphics[scale=0.5]{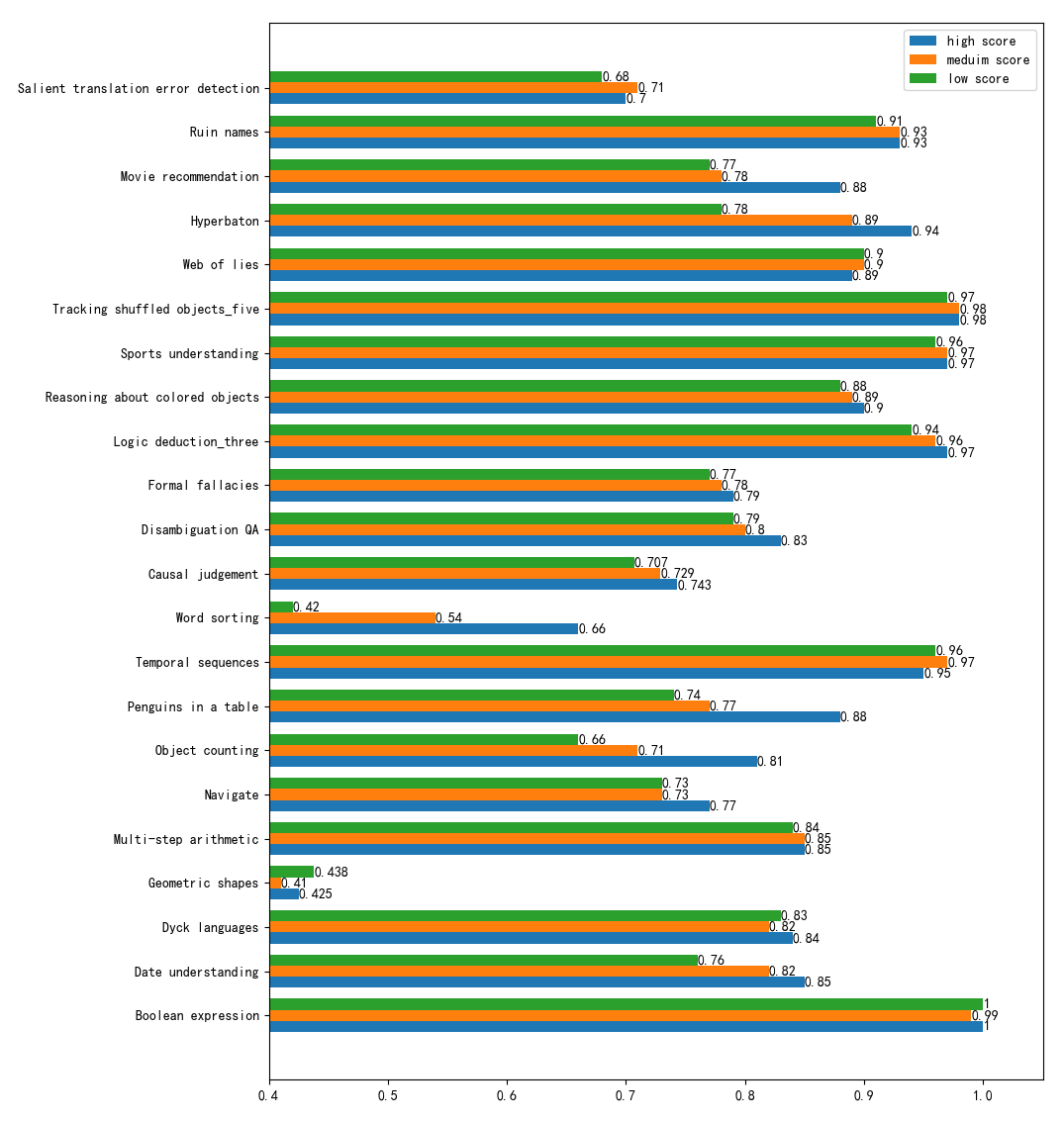}
	\caption{Results of correlation between the score and the test accuracy. Due to the high cost, We choose three scores:min, medium and max score of each individual task to evaluate on the test set.}

 \label{fig:score_bar}
 
\end{figure*}

\clearpage

\end{document}